\documentclass[pdflatex,sn-mathphys-num]{sn-jnl}

\usepackage{graphicx}%
\usepackage{multirow}%
\usepackage{amsmath,amssymb,amsfonts}%
\usepackage{amsthm}%
\usepackage{mathrsfs}%
\usepackage[title]{appendix}%
\usepackage{xcolor}%
\usepackage{textcomp}%
\usepackage{manyfoot}%
\usepackage{booktabs}%
\usepackage{algorithm}%
\usepackage{algorithmicx}%
\usepackage{algpseudocode}%
\usepackage{listings}%
\usepackage{booktabs} % 
\usepackage{hhline} % 
\usepackage{array}
\usepackage{array, tikz}

\theoremstyle{thmstyleone}%
\theoremstyle{thmstyletwo}%

\theoremstyle{thmstylethree}%

\begin{document}

\title[Article Title]{\bf Atom Cavity Encoding for NP-Complete Problems}

\author[1,2]{\fnm{Meng} \sur{Ye}}

\author*[1,2,3,4,5]{\fnm{Xiaopeng} \sur{Li}}\email{xiaopeng\underline{ }li@fudan.edu.cn}

\affil*[1]{State Key Laboratory of Surface Physics, and Department of Physics, Fudan University, Shanghai 200433, China}

\affil[2]{Shanghai Qi Zhi Institute, AI Tower, Xuhui District, Shanghai 200232, China}
\affil[3]{Hefei National Laboratory, Hefei 230088, China}

\affil[4]{Shanghai Artificial Intelligence Laboratory, Shanghai 200232, China}

\affil[5]{Shanghai Research Center for Quantum Sciences, Shanghai 201315, China}

\abstract{
We consider an atom-cavity system having long-range atomic interactions mediated by cavity modes. It has been shown that quantum simulations of spin models with this system can naturally be used to solve number partition problems. Here, we present encoding schemes for numerous NP-complete problems, encompassing the majority of Karp’s 21 NP-complete problems. We find a number of such computation problems can be encoded by the atom-cavity system at a linear cost of atom number. There are still certain problems that cannot be encoded by the atom-cavity as efficiently, such as quadratic unconstrained binary optimization (QUBO),  and the Hamiltonian cycle. For these problems, we provide encoding schemes with a quadratic or quartic cost in the atom number.  
We expect this work to provide important guidance to search for the practical quantum advantage of the atom-cavity system in solving NP-complete problems. 
Moreover, the encoding schemes we develop here may also be adopted in other optical systems for solving NP-complete problems, where a similar form of Mattis-type spin glass Hamiltonian as in the atom-cavity system can be implemented. 
 
}

\keywords{combinatorial optimization, quantum computation, cold atoms, cavity, quantum simulations}

\maketitle

\section{Introduction}\label{sec1}

Combinatorial optimization aims at finding an optimal solution from a large set of discrete solution space \cite{korte2011combinatorial,papadimitriou1998combinatorial}, which has important applications in diverse fields of technology and industry, including protein folding \cite{robert2021resource}, simulating molecular dynamics \cite{irie2021hybrid}, and modeling wireless communication networks \cite{durrquantum}. The optimization problems typically have an exponentially large solution space, and finding their optimal solution is NP-hard with classical computing. Quantum computing provides an alternative way to solve combinatorial optimization problems owing to its intrinsic quantum superposition parallelism. 
There have been several quantum algorithms developed for quantum optimization, including  quantum annealing~\cite{farhi2001quantum,farhi2002quantum,farhi2014quantum}, quantum approximate optimization algorithm (QAOA)~\cite{farhi2014quantum}, 
and quantum enhanced Markov chain (QEMC) \cite{layden2023quantum}. 
The study of practical quantum speedup with these quantum algorithms has attracted widespread attention in the last decade~\cite{albash2018adiabatic,hastings2021power,blekos2024review, luo2024adaptive}. 
%With the rapid progress in quantum optimization algorithms, encoding solutions of NP-complete problems as ground states of Hamiltonians has gained significant traction, particularly in near-term quantum computing architectures such as superconducting quantum processors, Rydberg atom arrays, and atom-cavity systems. 

Recent research has shown significant advancements in encoding solutions for NP-complete problems as ground states of Hamiltonians, particularly in near-term quantum computing platforms such as superconducting quantum processors, Rydberg atom arrays, and atom-cavity system discussed in this work. Among these quantum computing platforms, superconducting processors such as D-Wave systems utilize the Chimera or Pegasus graph structure representing interactions between qubits. In D-Wave systems, there exists a significant challenge: the embedding cost for arbitrary all-to-all connectivity problems scales quadratically with the number of logical qubits, demanding substantial resources~\cite{date2019efficiently, boothby2020next}. In Rydberg
atom arrays, the unique Rydberg blockade mechanism enables a natural encoding of maximum-weight independent set (MWIS) problems~\cite{nguyen2023quantum}. Due to the geometric limitations arising from the interaction range of Rydberg blockade, encoding the maximum weighted independent set problems still require a quadratic scaling in the number of physical qubits, even when advanced techniques such as parity encoding or quantum wiring are employed~\cite{qiu2020programmable,kim2022rydberg,lanthaler2023rydberg,goswami2024solving}.

Our proposed atom-cavity encoding scheme offers a distinct advantage. It exhibits all-to-all couplings between any two qubits, which can naturally encode number partition problems~\cite{ye2023universal, anikeeva2021number, ramette2022any, cesa2023universal}. Additionally, for other problems such as 3-SAT and vertex cover, the atom-cavity system also exhibits a linear cost scaling with respect to the number of qubits, marking a scaling improvement over Rydberg atom arrays and D-Wave systems. Despite these benefits, the efficient encoding of other NP-complete problems in the atom-cavity system remains an open challenge.

In this paper, we examine numerous NP-complete problems, for which the encoding schemes are provided considering the atom-cavity system. For a number of graph, assignment, and set-partition problems, we show the atom number cost has a linear scaling, which indicates the atom-cavity system is more efficient than the Rydberg system in solving NP-complete problems in terms of the atom number cost. 
For other problems such as QUBO, integer programming, Hamiltonian cycle, we find a heavier overhead in the encoding. The results are summarized in Table~\ref{tab: main results}. 
Our theory implies the atom-cavity system is a promising platform to achieve a practical quantum advantage in solving combinatorial optimization problems.

\begin{table}[!h]

\caption{Problems encoded in atom-cavity system}
\label{tab: main results}
\begin{tabular}{|m{1.9cm}| m{1.9cm}| m{1.9cm}|| m{1.9cm}|| m{1.9cm}|}
\hline
\multicolumn{3}{|c||}{\textbf{Linear Scaling}} & \multicolumn{1}{ | c ||}{\textbf{Quadratic Scaling}}&\multicolumn{1}{ | c |}{\textbf{Quartic Scaling}}\\ %\hhline{|=|=|=||=||=|}
\hhline{|===::=::=|}
\textit{Sets and Partitions} & \textit{Graph problems}
     & \textit{Assignment problems} & \begin{tikzpicture}[scale=1]
             \draw[dashed] (1,1.5) -- (2,1.5);
              \end{tikzpicture}& \begin{tikzpicture}[scale=1]
              \draw[dashed] (1,1.5) -- (2.3,1.5);
              \end{tikzpicture} \\

\hhline{|t:=t:=t:=t:=t:=t:|}
Subset Sum Problem & Maximum Independent Set & 3-SAT & QUBO & Job Sequencing Problem  \\ \hhline{|-|-|-||-||-|}
Exact Cover & Vertex Cover & 3-Coloring & Binary Integer Linear Programming  & Directed Hamiltonian Cycle \\ \hhline{|-|-|-||-||-|}
    Maxcut & Clique & Dominatinig Set Problem & Knapsack with Integer Weights & Traveling Salesman \\ \hhline{|-|-|-||-||-|}
    Set Packing& Matching &    &   &    \\ \hline
\end{tabular}

\end{table}

\section{Background}\label{sec2}
In this section, we introduce the experimental system and its corresponding Hamiltonian. Subsequently, we commence with the encoding of the subset sum problem, serving as our theoretical foundation before extending it to other NP-complete problems. In the final subsection, a brief experimental implementation of our encoding scheme is presented.

\subsection{Cold Atoms in Cavity}

This work is initially inspired by a recent breakthrough in atom-cavity system, where a universal quantum optimization architecture was developed \cite{ye2023universal}. In this system, deterministically assembled atom arrays are coherently manipulated with optical pulses and entangled through a four-photon process resulting in long-range interactions that naturally encode number partition problems. In this work, we will first utilize such a system to encode subset sum problems, thereby serving as a foundational step towards extending our approach to encompass a broader range of NP-complete problems.

Now we consider the physical platform more specifically. Suppose $N$ three-level atoms in a high-finesse
optical cavity and each atom realizes a qubit with two internal ground state with an intrinsic energy splitting
$\hbar \Delta_F$ representing $|0\rangle$ and $|1\rangle$. They are both coupled to the excited state $|e\rangle$ by the same cavity mode with detunings $\Delta+\Delta_F$ and $\Delta$, respectively. After rotating wave approximation and adiabatically
eliminating the excited state, the quantum dynamics of the atoms mediated by cavity mode can be described by a Hamiltonian 

\begin{equation}
\hat H_{\text{eff}}/\hbar = \sum_{j=1}^{N} \frac{ \delta_m}{2} \hat \sigma_{i}^{z} + g_4\left[\sum_{i=1}^{N}\lambda_i\hat \sigma_i^x\right]^2.
\end{equation}

Here $\lambda_i=\sin(QX_i)$, where Q denotes the wave-vector of the cavity mode. Therefore, the interactions could be programmed by manipulating atom positions. $g_4$ is a programmable parameter that is proportional to the square of Rabi frequency $\Omega$. $\delta_m$ is another dynamically tunable parameter that is related to Rabi frequency and laser detuning.

In addition to the four-photon process described, the programmable Rabi transitions between $|0\rangle$ and $|1\rangle$ also exist which can be induced by interactions between laser and atoms. Considering these processes, the collective Hamiltonian governing the atom-cavity system can be expressed as follows

\begin{equation}
    H = \sum_{j=1}^{N} \frac{ \delta_m}{2} \hat \sigma_{i}^{z}+\sum_{j=1}^{N} B_i \hat \sigma_{i}^{x} + g_4\left[\sum_{i=1}^{N}\lambda_i\hat \sigma_i^x\right]^2,
    \label{eq:cavityeff} 
\end{equation}
where $\delta_m$, $B_i$, $\lambda_i$ are all programmable parameters that could help with encoding for target Hamiltonian.

\subsection{Encoding for Subset Sum Problem }

Subset sum problem is a decision problem to determine whether there exists a subset of a given set of numbers whose elements sum up to a specified target integer. Formally, given a finite set of integers $S=\{a_{1}, a_{2},\dots, a_{n}\}$ and a target $T$, the question is whether a non-empty subset $S'\in S$ such that the sum of its elements is equal to the target $T$
\begin{equation}
    \sum_{a_{i}\in S'} a_{i}=T.
\end{equation}
The classical energy function can be represented as
\begin{equation}
    E=(\sum_{i}x_{i}a_{i}-T)^2,
\end{equation}
where $x_{i}=1$(or $0$) if the number $a_{i}$ is in (or not in) subset $S'$.
Generally, the problem Hamiltonian can be written as
\begin{equation}
    H= \sum_{i}h_{i}\hat \sigma_{i}^{x}+\sum_{i,j}\lambda_{i}\lambda_{j}\hat \sigma_{i}^{x}\hat \sigma_{j}^{x},
\end{equation}
where $\lambda_i=a_i/a_{max}$, $h_i=-2Ta_i/a_{max}^2$. Upon comparison with the Hamiltonian of atoms in cavity Eq.(\ref{eq:cavityeff}), it becomes evident that the problem formulation can be naturally encoded within this physical framework. 

The discussion above serves as an essential starting point for the majority of encoding schemes presented in this article. Instead of directly encoding the computational problem, an approach is adopted in which the subset sum problem is regarded as an intermediate bridge to a practical atomic system.

The subsequent sections focus exclusively on the atom-cavity system, it is noteworthy that our encoding scheme can also be applied to a spatial-photonic Ising machine, which possesses a similar Hamiltonian structure~\cite{pierangeli2019large}. In this system, effective spin variables were established to correspond to spatial points within the optical field, resulting in a spatial light intensity that takes the form of an Ising Hamiltonian
\begin{equation}
    H = -\sum_{ij}J_{ij}\sigma_i\sigma_j,
\end{equation}
with spin interactions 
\begin{equation}
    J_{ij}=\xi_i\xi_j\widetilde{I}(i,j).
\end{equation}
Here, $\xi_i$ denotes the field amplitude incident on pixel $i$ independently, analogous to $\lambda_i$, which is solely related to the positions of the atoms in our atom-cavity system. Therefore, the encoding scheme developed for the atom-cavity system in our study is also applicable to other optical systems.

\subsection{Quantum Optimization Algorithms}

Quantum optimization aims to utilize
quantum fluctuations to solve difficult binary optimization problems \cite{papadimitriou2013combinatorial}. Over the past two decades, various quantum algorithms, including quantum approximate optimization algorithms (QAOA) and adiabatic quantum computing (AQC), have been proposed.  

The quantum approximate optimization algorithm (QAOA) is a hybrid quantum-classical algorithm that iteratively applies a series of quantum gates to an initial quantum state \cite{farhi2014quantum}. Increasing the number of iterative layers in QAOA can degrade performance due to decoherence in current NISQ devices and the barren plateau issue in classical optimization \cite{cerezo2021variational}. Therefore, a number of ansatz variants have been proposed recently \cite{blekos2024review}, including adaptive bias QAOA \cite{yu2022quantum,luo2024adaptive}, recursive QAOA \cite{bravyi2020obstacles}, and warm-starting QAOA \cite{egger2021warm}.

The adiabatic quantum computing (AQC) is a quantum computing paradigm based on the adiabatic theorem. A computational problem is encoded into a problem Hamiltonian, with its ground states representing the solutions. The system begins in the ground state of an initial Hamiltonian and evolves slowly towards the final Hamiltonian. Beyond the simplest linear evolution path, various techniques have been developed in the past decade to enhance the success probability of AQC. One approach involves optimizing the adiabatic evolution path \cite{albash2018adiabatic, guery2019shortcuts, hegade2021shortcuts, schiffer2022adiabatic}, and the design of the path configuration could be enhanced through the application of machine learning techniques \cite{liu2021rigorous,lin2020quantum,lin2022hard,chen2022optimizing}. Additionally, iterative reverse annealing schemes have also been applied to further improve the performance of AQC \cite{perdomo2011study,chancellor2017modernizing,yamashiro2019dynamics}.

Here, we present an experimental setup within the framework of adiabatic quantum computing (AQC) as an illustrative example. We revisit the physical Hamiltonian of the atom-cavity system, previously derived as Eq.(\ref{eq:cavityeff})
\begin{equation*}
    H = \sum_{j=1}^{N} \frac{ \delta_m}{2} \hat \sigma_{i}^{z}+\sum_{j=1}^{N} B_i \hat \sigma_{i}^{x} + g_4\left[\sum_{i=1}^{N}\lambda_i\hat \sigma_i^x\right]^2,
\end{equation*}
 where $B_i$, $\delta_m$ and $g_4$ are dynamically tunable in experiments. we could translate the evolution path into these parameters. At the beginning of adiabatic evolution, the global transverse-field parameter $\delta_m$ is initialized with a finite value, while $B_i$ and $g_4$ are both set to zero at $t=0$. Then, during the adiabatic evolution process, we tune the detuning as $\delta_m(t) = [1-s(t)]\delta_m(0)$. Simultaneously, we ramp the intensity of the operating lasers and Rabi lasers to the desired value respectively, so that $g_4(t) = s(t)g_4$ and $B_i(t)=s(t)B_i$. According to the adiabatic theorem \cite{born1928beweis}, the solutions can be obtained by measuring the atomic spins in the Pauli-$\hat \sigma_x$ basis.
 
%{\red [Comment: Add discussion about optimizing adiabatic path. Cite Jian's paper, and also others like Cirac PRX Quantum, Xi Chen ShortCut, Chang-Yu Hsieh Nature Machine intelligience. ---XL ]}

\section{Problems with Linear Overhead}\label{sec3}
In this section, we will introduce problems that can be encoded into an atom-cavity system with linear overhead. For each problem in this section, we will give a corresponding subset sum formula such that the subset sum could be satisfied if and only if the original problem has solutions. Therefore, once the subset sum problem is reduced from its original form, we could proceed to encode these problems into an atom-cavity system, aligning with the subset sum problem formulation as explained in Section 2.
\subsection{3-SAT}
3-SAT is one of Karp's 21 NP-complete problems, and it is always used as a starting point for proving that other problems are also NP-hard \cite{karp2010reducibility}, which has wide application in complexity theory \cite{aho1974design}, cryptography \cite{massacci2000logical} and artificial intelligence \cite{vizel2015boolean}. In 3-SAT, the input is a Boolean formula expressed as a conjunction of clauses $\phi=(C_{1}\land C_{2}\dots \land C_{m})$, where each clause, $C_{i}=(x_{i1}\oplus B_{i1})\lor(x_{i2}\oplus B_{i2})\lor(x_{i1}\oplus B_{i1})$, consists of exactly three literals (variables or their negations) connected by logical OR operators. The objective is to ascertain whether there exists an assignment {$x_{1},x_{2},\dots,x_{N}$} of truth values (true or false) to the variables that satisfy the entire formula, making each clause true simultaneously. 

As explained in the pre-section, we will first reduce this 3-SAT problem to a subset sum problem, which has been discussed in \cite{ye2023universal}. We introduce

\begin{align}
    a_{i}&=\sqrt{\alpha_{m+i}}\enspace+\sum_{j:x_{i} {\rm in} c_{j}}^{m}\sqrt{\alpha_{j}} , \\
     b_{i}&=\sqrt{\alpha_{m+i}}\enspace+\sum_{j:\overline{x_{i}} {\rm in} c_{j}}^{m}\sqrt{\alpha_{j}},   \\ 
     c_j& = d_j = \sqrt{\alpha_j}, 
\end{align}
with $\alpha_p$ the $p$-th squarefree integer, starting from $1$. Here, we have  $i\in [1, n]$, and $j \in [1, m]$. The numbers $\{a_i\}$, $\{b_i\}$, $\{c_j\}$, and $\{d_j\}$ form a set ${\cal R}$, which then has $2n+2m$ number elements  to be referred to  as $r_k (I, B) $, with $k \in [1, 2n+2m]$. The target sum is

\begin{equation}
    T =\sum_{i=1}^{n}\sqrt{\alpha_{m+i}}+3\sum_{j=1}^{m}\sqrt{\alpha_{j}}.
\end{equation}
 We will prove that solving the 3-SAT problem is equivalent to solving the constructed subset sum problem $\sum_k y_k r_k (I, B) = T$.

 First, we shall prove that if {$x_{i}$} is a solution for the 3-SAT clause $\phi$, then there is a corresponding solution {$y_{k}$} satisfying the equation $\sum_k y_k r_k (I, B) = T$. We can determine {$y_{k}$} as like $y_{k\in[1,n]}=x_k$ and $y_{k\in[n+1,2n]}=\bar x_{k-n}$. Thus we can get

\begin{equation}
    \sum_{k=1}^{2n} y_k r_k (I, B) = \sum_i x_i a_i + \bar x_i b_i = \sum_{i=1}\sqrt{\alpha_{m+i}}+\sum_{j=1}\sqrt{\alpha_{j}}\lbrack \sum_{i:x_i\in C_j}x_i+\sum_{i:\bar x_i\in C_j}\bar x_i \rbrack.
\end{equation}
 Since {$x_i$} is the solution of 3-SAT problem, then the summation $g_i=\sum_{i:x_i\in C_j}x_i+\sum_{i:\bar x_i\in C_j}\bar x_i$ takes value of 1, 2 or 3. For the three cases with $l_j=1$, 2 and 3, we choose $(y_{2n+j}=1,y_{2n+m+j}=1)$, $(y_{2n+j}=1,y_{2n+m+j}=0)$, and $(y_{2n+j}=0,y_{2n+m+j}=0)$, respectively. For all three cases, we have 
 \begin{equation}
     \sum_{k=1}^{2n+2m} y_k r_k (I, B) = T.
 \end{equation}

Conversely, if {$y_k$} is the solution for the subset sum problem. Then we can quickly find out 

\begin{align}
    & y_i+y_{n+i}=1, \text{ for } i\in[1,n],\\
    & \lbrack \sum_{i:x_i\in C_j}y_i+\sum_{i:\bar x_i\in C_j} y_{n+i} \rbrack+y_{2n+j}+y_{2n+m+j}=3,
\end{align}
for the linear independence of radicals obeyed by the squarefree integers. It is then guaranteed that $y_i=\bar y_{n+i}$. And we here choose $x_i=y_i$, then it becomes
\begin{equation}
    \lbrack \sum_{i:x_i\in C_j}x_i+\sum_{i:\bar x_i\in C_j} \bar x_{n+i} \rbrack+y_{2n+j}+y_{2n+m+j}=3.
\end{equation}
Take the values of $y_{2n+j}+y_{2n+m+j}$ into consideration,
\begin{equation}
    \lbrack \sum_{i:x_i\in C_j}x_i+\sum_{i:\bar x_i\in C_j} \bar x_{n+i} \rbrack = 1,2, \text{ or }3, \text{ for all }j,
\end{equation}
which indicates the 3-SAT problem is resolved. Meanwhile, the atom number of encoding the 3-SAT problem is linear,
\begin{equation}
   \text{Encoding cost}= 2(n+m).
\end{equation}

\subsection{Vertex Cover}

Given an undirected graph $G=(V, E)$, where $V$ is the set of vertices and $E$ is the set of edges, a vertex cover $V'$ of $G$ is a subset of $V$ such that for every edge $(u,v)$ in $E$, at least one of the endpoints $u$ or $v$ belongs to $V'$. The vertex cover problem is to decide if a given graph $G=(V, E)$ has a vertex of size $k$. It serves as a paradigm for numerous real-world and theoretical challenges, particularly within the domain of metabolic engineering as evidenced by recent research endeavors \cite{hossain2020automated,reis2019simultaneous}. We reduce it to a number partition problem as discussed in \cite{ye2023universal}.

%A vertex cover of a graph is a set of vertices that touches every edge of the graph. The vertex cover problem is to decide if a given graph $G=(V, E)$ has a vertex of size $k$. It serves as a paradigm for numerous real-world and theoretical challenges, particularly within the domain of metabolic engineering as evidenced by recent research endeavors\cite{hossain2020automated,reis2019simultaneous}. we reduce it to a number partition problem first.

We consider a graph $G=(V,E)$ with $n$ vertices, $v_{1},\dots, v_{n}$, and $m$ edges, $e_{1},\dots, e_{m}$. We introduce $n$ numbers corresponding to the vertices,
\begin{equation}
    a_{i}=\sqrt{\alpha_{m+1}}+\sum_{l:i\in{e_{l}}}\sqrt{\alpha_{l}},\  i=1,\dots,n,
\end{equation}
and $m$ numbers corresponding to the edges,
\begin{equation}
    b_{j}=\sqrt{\alpha_{j}},\ j=1,\dots,m.
\end{equation}
These numbers ($a_i$ and $b_j$) form a set $R$ with size $n+m$. Thus in the fundamental encoding logic, solving the vertex cover problem is equivalent to solving $\sum_{k=1}^{n+m}y_k r_k = T$, where
\begin{equation}
    T=k\cdot\sqrt{\alpha_{m+1}}+\sum_{i=1}^{m}2\cdot\sqrt{\alpha_{i}}.
\end{equation}
The proof of their equivalence is similar
to the 3-SAT encoding. We first explain that if $G$ has a vertex cover of size $k$, then we will find the corresponding {$y_k$}.

Suppose that graph has a vertex cover $U$ of size $k$, then for each vertex $v_i$, if $v_i \in U$, then $y_i=1$, and if $v_i \notin U$, then we choose $y_i=0$. The summation for the first $n$ number is
\begin{equation}
    k\cdot\sqrt{\alpha_{m+1}}+\sum_{j=1}^{m}g_{j}\cdot\sqrt{\alpha_{i}},
\end{equation}
 where $g_{j}$ is the number of endpoints in edge $e_{i}$ that belong to vertex cover and equals 1 or 2. We can complete our reduction process by selecting $y_{n+j}=1$ for the case when $g_{j}=1$, and $y_{n+j}=0$ for the case when $g_{j}=2$.

Conversely, if we have the solution for $\sum_{k=1}^{n+m}y_k r_k = T$. It can be readily observed that the term $k\cdot\sqrt{\alpha_{m+1}}$ arises from the $k$ values of ${a_{i}}$. Subsequently, we take into account the factors of $\sqrt{\alpha_{i}}$, where $i\in[1,m]$. Here, it is important to note that the contributions to the target sum from the elements of ${b_{i}}$ are limited to at most one $\sqrt{\alpha_{i}}$. Consequently, it can be inferred that the $k$ values of ${a_{i}}$ contain all of the $\sqrt{\alpha_{i}}$ that correspond to the vertices that form a vertex cover for the given graph $G$.

We finished the encoding process for the vertex cover problem, and the atom number consumption is 
\begin{equation}
   \text{Encoding cost}= n+m.
\end{equation}

\subsection{Maximum Independent Set}
Given an undirected graph $G=(V, E)$, an independent set is a subset of vertices in which no two vertices are adjacent. The MIS problem is to decide if the given graph has an independent set of size $k$. MIS is an NP-complete problem with applications in many areas, including network design, social network analysis, scheduling, and operations research. Notably, the maximum independent set (MIS) problem is a fundamental combinatorial optimization problem in graph theory and has extensive research results in Rydberg atom system \cite{pichler2018quantum,byun2022finding,nguyen2023quantum,lanthaler2023rydberg}.

According to definitions of the MIS problem and vertex cover problem, it is evident that if a graph $G=(V, E)$ has a vertex cover $U$ then the $V/U$ constitutes an independent set. Hence, the encoding methodology established for resolving the vertex cover problem serves as a direct inspiration for formulating an encoding strategy to address the MIS problem.

Considering a graph $G=(V,E)$ with $n$ vertices $v_{1},v_{2},\dots,v_{n}$ and $m$ edges $e_{1},e_{2},\dots,e_{m}$. Similarly, we introduce $n$ numbers corresponding to vertices and $m$ numbers corresponding to edges,
\begin{align}
    a_{j}&=\sqrt{\alpha_{m+1}}+\sum_{l:j\in{e_{l}}}\sqrt{\alpha_{l}},j=1,\dots,n, \\
 b_{i}&=\sqrt{\alpha_{i}}, i=1,\dots,m,
\end{align}
and set the target sum as
\begin{equation}
T=k\cdot\sqrt{\alpha_{m+1}}+\sum_{i=1}^{m}\sqrt{\alpha_{i}}.
\end{equation}

The equivalence proof between subset sum and MIS is analogous to the vertex cover encoding previously discussed. If graph $G$ has an independent set of size $k$, then the sum associated with the vertices in this independent set is given by $k\cdot\sqrt{\alpha_{m+1}}+\sum_{i=1}^{m}g_{i}\cdot\sqrt{\alpha_{i}}$, where $g_{i}$ is either 0 or 1, subject to the independence constraint. Similar to the vertex cover problem, the rest of target $T$ can be completed by adding $b_{i}$ according to the value of $g_{i}$. Conversely, given a solution to the subset sum problem, it becomes obvious that the term $k\cdot\sqrt{\alpha_{m+1}}$ must originate from a selection of $k$ values of ${a_{i}}$. Accordingly, we select the corresponding vertices {$v_i$} to form a set. As implied by the $\sum_{i=1}^{m}\sqrt{\alpha_{i}}$ term, we ensure that the chosen vertices do not share any common edges. Therefore, the proof has been finished, and the encoding overhead is the same as the vertex cover, 
\begin{equation}
    \text{Encoding cost}= n+m.
\end{equation}

\subsection{Clique}

A clique $V'$ of a given graph $G=(V, E)$ is a subset of vetices such that $\forall u,v\in V'$, where $u\neq v$, we have $(u,v)\in E$. Therefore, the clique problem is to determine whether a clique of specified size $k$ exists for a given graph, which has been proved to be an NP-complete problem \cite{karp2010reducibility}.

%A clique is defined as a subset of vertices within the graph, where every pair of distinct vertices is connected by an edge, forming a fully connected substructure. Therefore, the clique problem is to determine whether a clique of specified size $k$ exists for a given graph, which has been proved to be an NP-complete problem.

For a given graph $G=(V, E)$, we will construct a corresponding number set as follows. We first introduce $n$ numbers for vertices
\begin{equation}
    a_{i}=\sqrt{\alpha_{n+1}}+(k-1)\cdot\sqrt{\alpha_{i}},\ i=1,2\dots,n.
\end{equation}
For edge between vertex $i$ and $j$, we introduce the corresponding number
\begin{equation}
    b_{\langle i,j\rangle}=2\sqrt{\alpha_{n+1}}-\sqrt{\alpha_{i}}-\sqrt{\alpha_{j}}.
\end{equation}
Finally, we can set the target as
\begin{equation}
    T=k^{2}\sqrt{\alpha_{n+1}}.
\end{equation}

Now we explain the equivalence between the subset sum problem and the original clique problem. First, if graph $G$ has a clique $V'$ of size $k$, the construction of the associated subset proceeds as follows. We put $a_i$ into the corresponding subset if vertices $v_i\in V'$. For each pair $ v_i,v_j\in V'$, add $b_{\langle i,j\rangle}$ into the subset. Then the summation can be worked out,
\begin{equation}
    k\sqrt{\alpha_{n+1}}+\frac{k(k-1)}{2}\cdot 2\sqrt{\alpha_{n+1}}=k^2\sqrt{\alpha_{n+1}}.
\end{equation}

Conversely, the existence of a clique of size $k$ in the graph $G$ can be inferred from the presence of a solution to the subset sum problem. Now we look back at the subset problem. First, we construct a subset $S$ with $p$ numbers of the type associated with vertices, denote as $\{a_{s(1)}, \dots, a_{s(i)}, \dots, a_{s(p)}\} $. The summation over this subset takes the form $p\sqrt{\alpha_{n+1}}+(k-1)\sum_{i=1}^{p}\sqrt{\alpha_{s(i)}}$. Since the target sum only contains $\sqrt{\alpha_{n+1}}$, terms $\sqrt{\alpha_{i\in[1,n]}}$ must be eliminated by incorporating numbers of the type associated with edges into the subset.  Assume that for any pair $(a_{s(i)}$, $a_{s(j)})$ in the subset, the number $b_{\langle s(i),s(j)\rangle}$ exists and is added to the subset. Then the summation transforms to $p^2\sqrt{\alpha_{n+1}}+(k-p)\sum_{i=1}^{p}\sqrt{\alpha_{s(i)}}$. The target sum can be achieved when $p \geq k$. Specifically, if $p=k$, the summation precisely matches the target sum $T=k^2\sqrt{\alpha_{n+1}}$. If $p > k$, a subset of $\{a_{s(i)}\}$ of size $k$ can be selected and complemented with $b_{\langle s(i),s(j)\rangle}$ for any pair $(a_{s(i)}$, $ a_{s(j)})$. The discussion above implies that at least the corresponding graph $G$ has a clique of size $k$. Finally, we can construct the clique by selecting vertices $v_i$ whenever $a_i$ constitutes a solution for the subset sum problem.

%The construction of the clique follows that We put together the vertices $v_i$ into the vertex set whenever $a_i$ constitutes a solution for the subset sum problem. 

%For the original subset sum problem, because the target sum only contains $\sqrt{\alpha_{n+1}}$, the terms $\sqrt{\alpha_{i\in[1,n]}}$ require elimination through the formation of a clique. Specifically, each vertex $v_i$ is interconnected with all other vertices, resulting in a net contribution of $(k-1)\cdot\sqrt{\alpha_{i}}$ for each vertex, while edges $\langle i,j\rangle$ associated with the vertex $v_i$ contribute $-(k-1)\cdot\sqrt{\alpha_{i}}$. The vanishing of $\sqrt{\alpha_{i\in[1,n]}}$ imposes constraints on the vertex set, forcing it to form a clique.

Here we complete the proof, and the encoding cost is the same as vertex cover and MIS
\begin{equation}
   \text{Encoding cost}= n+m.
\end{equation}

\subsection{Maximum Matching}

Given an undirected graph $G=(V, E)$, a matching in $G$ is a subset $M$ of $E$ such that for any two distinct edges $(u_1, v_1)$ and $(u_2, v_2)$ in $M$, satisfying $u_1 \neq u_2$, $u_1 \neq v_2$, $v_1 \neq u_2$, and $v_1 \neq v_2$ – i.e., no two edges in the matching share common vertices. A maximum matching is a maximal matching that contains the largest number of edges. Finding a maximum matching is NP-hard, while its corresponding decision problem is NP-complete.
%Given a graph $G=(V, E)$, a matching $M$ in $G$ is defined as a set comprising pairwise non-adjacent edges, with the additional constraint that none of the edges within $M$ form loops. Formally, this implies that no two edges in the matching share common vertices. Then a maximal matching is a matching $M$ of graph $G$ that is not a subset of any other matching, and a maximum matching is a maximal matching that contains the largest number of edges. Finding a maximum matching is NP-hard, while its corresponding decision problem is NP-complete.

To encode the matching problem to the atom-cavity system, we reduce it to a subset sum problem. Given graph $G$, with vertices $\{v_{1},v_{2},\dots,v_{n}\}$ and edges $\{e_{ij}\}$. We can introduce $n$ numbers associated with vertices and $m$ numbers associated with edges,
\begin{align}
    a_{i}&=\sqrt{\alpha_{i}},\text{ }i=1,2,\dots,n ,\\
    b_{\langle i,j \rangle}&=\sqrt{\alpha_{n+1}}+\sqrt{\alpha_{i}}+\sqrt{\alpha_{j}}.
\end{align}
The corresponding target can be set as
\begin{equation}
T=k\cdot\sqrt{\alpha_{n+1}}+\sum_{i=1}^{n}\sqrt{\alpha_{i}}.
\end{equation}

The proof logic is the same as encoding for the maximum independent set problem. And the encoding cost is also the same
\begin{equation}
   \text{Encoding cost}= n+m.
\end{equation}

\subsection{Exact Cover}
%The exact Cover Problem seeks to identify a subset of a given collection of sets such that each element appears exactly once in the chosen subsets, and every element in the universal set is covered exactly once. In essence, the objective is to find a precise cover, leaving no element uncovered or duplicated.
$X$ is a finite set $\{I_j\}$ of size $m$. Let $S$ be a collection of subsets of $X$, and whose size is $n$. The exact cover problem involves determining whether there exists a subcollection $S'$ such that each element in $X$ is covered exactly once by the union of sets in $S'$. Furthermore, each set in $S'$ must be distinct, ensuring that the cover is exact and non-redundant. The exact cover problem is a classical NP-complete problem, signifying its computational complexity and its relevance in the study of algorithmic efficiency \cite{korte2011combinatorial}.

For each subset in $S$, we can introduce a number
\begin{equation}
    a_{i}=\sum_{I_j\in S_{i}}\sqrt{\alpha_{j}},
\end{equation}
and set the target for the number set
\begin{equation}
    T=\sum_{i=1}^{m}\sqrt{\alpha_{i}}.
\end{equation}

If an exact cover exists, denoted by a subset of ${a_i}$, then following the previously assumed conditions, each element $I_j$ is associated with $\sqrt{\alpha_{j}}$. An exact cover comprises each element exactly once, resulting in a summation of $\sum_{i=1}^{n}\sqrt{\alpha_{i}}$. Conversely, the existence of a solution to the subset sum problem implies that a subset $S'$ covers all the elements without repetition or omission. Finally, we get the cost of atoms,
\begin{equation}
    \text{Encoding cost}= n.
\end{equation}

\subsection{Set Packing}

$X$ is a finite set, and let $S=\lbrace S_1, S_2,\dots, S_m\rbrace$ be a collection of subsets of $X$. The set packing problem is to determine if there exists $k$ subsets $S_{\alpha1},\dots, S_{\alpha k}$ in $S$ satisfying $S_{\alpha i}\cap S_{\alpha j}=\emptyset$ for all distinct $S_{\alpha i}$, $S_{\alpha j}$ in such $k$ subsets.

For each element in set X, we introduce
\begin{equation}
    b_i=\sqrt{\alpha_i},\quad i=1,2\dots,n.
\end{equation}
For each subset in S, we introduce
\begin{equation}
    a_{i}=\sqrt{\alpha_{n+1}}+\sum_{j\in S_{i}}\sqrt{\alpha_{j}}.
\end{equation}
Finally, set the target as
\begin{equation}
    T=k\sqrt{\alpha_{n+1}}+\sum_{i=1}^{n}\sqrt{\alpha_{i}}.
\end{equation}
The proof logic here has been discussed in the section for the maximum independent set problem. Moreover, the consumption of encoding atoms is also
\begin{equation}
    \text{Encoding cost}= n+m.
\end{equation}

\subsection{Maxcut}
Given an undirected graph $G=(V, E)$ with a set of vertices $v_i,\dots,v_n$ and a set of edges $e_1,\dots,e_m$, the maxcut problem is to find a partition $(U, W)$ of $V$ that maximizes the cut size, which is the number of edges $(u,v) \in E$ where $u \in U$ and $v \in W$. In deterministic language, it is an NP-complete problem which is to decide if the given graph $G$ has a partition whose cut size is $k$.

We introduce a set of numbers to reduce it as a subset sum problem. For each vertex $i$ we have
\begin{equation}
    a_i = \sum_{j:v_i\in e_j} \sqrt{\alpha_j}+d_i\sqrt{\alpha_{m+1}},
\end{equation}
where $d_i$ is the degree of vertex $v_i$. For each edge $e_j$ we have
\begin{equation}
    b_j= -2\sqrt{\alpha_j}-2\sqrt{\alpha_{m+1}}.
\end{equation}
To complete the encoding process, for each edge $e_j$, we additionally include 
\begin{equation}
    c_j = -\sqrt{\alpha_j}.
\end{equation}
The corresponding target sum is 
\begin{equation}
    T = k\sqrt{\alpha_{m+1}}.
\end{equation}

We present a brief proof as follows. Initially, we demonstrate that if the given graph $G$ has a partition with cut size $k$, then the corresponding subset sum problem has a solution. Suppose the partition is $(U, W)$. We can construct the number subset based on $U$; similarly, constructing the subset based on $W$  also yields a valid approach. For each vertex $ v_i\in U$, we include the corresponding $a_i$ in the subset. The intermediate sum then becomes 
\begin{equation}
    \sum_{i:v_i\in U}\sum_{j:v_i\in e_j (v_i)}\sqrt{\alpha_j}+\sum_{i:v_i\in U}d_i\sqrt{\alpha_{m+1}}.
\end{equation}
Considering the target sum $T$, which contains only $\sqrt{\alpha_{n+1}}$. Then we have to select numbers from $\{b_j$\} and $\{c_j\}$ to match the target sum according to the following rules: if one vertex of edge $e_j$ is in $U$, include $c_j$ to the number subset; if both vertex of edge $e_j$ are in $U$, include $b_j$ to the number subset. This adjustment ensures that the coefficient of $\sqrt{\alpha_{n+1}}$ becomes $k$, and eliminate all $\sqrt{\alpha_{j\in[1,m]}}$, thereby achieving the target sum.

Conversely, if the subset sum problem is solved, we can construct the graph partition based on its solution. Given that the target sum includes only $\sqrt{\alpha_{m+1}}$, we initially consider a number subset $S$ that includes $\{a_{s_1},\dots,a_{s_p}\}$. Summing over $S$ we obtain 
\begin{equation}
    \sum_{i=1}^{p}\sum_{j:v_{s_i}\in e_j} \sqrt{\alpha_j}+\sum_{i=1}^{p}d_{s_i}\sqrt{\alpha_{m+1}}.
\end{equation}
In this expression, the coefficient of each $\sqrt{\alpha_{j\in[1,m]}}$ is either 1 or 2. If the coefficient of $\sqrt{\alpha_{j}}$ is 1, we add $c_j$ to the subset $S$; if the coefficient of $\sqrt{\alpha_{j}}$ is 2, we add $b_j$ to the subset $S$. After this adjustment, the sum of $S$ becomes 
\begin{equation}
    \sum_{i=1}^{p}d_{s_i}\sqrt{\alpha_{m+1}}-2\sum_{(s_i,s_j)\in E}\sqrt{\alpha_{m+1}}.
\end{equation}
By examining this formulation, we can conclude that the coefficient of $\sqrt{\alpha_{m+1}}$ represents the cut size between $U$ and $V\setminus U$, where $U = \{ v_{s_1},\dots,v_{s_p}\}$ corresponds to the number subset $S$. Because the subset sum problem has a solution that achieves the target sum $k\sqrt{\alpha_{m+1}}$, the graph $G$ has a cut of size $k$, as previously discussed. The encoding cost is  
\begin{equation}
    \text{Encoding cost} = n+2m.
\end{equation}

%As is well-known in research, the maximum cut problem is equivalent to minimizing the ground energy of an Ising spin model\cite{barahona1988application}. Given a graph $G=(V,E)$ with associated edge weights {$w_{i,j}$}, the corresponding Hamiltonian can be represented as
%\begin{equation}
%    H=\sum_{\langle i,j \rangle \in E}-w_{i,j}(x_{i}+x_{j}-2x_{i}x_{j})
%\end{equation}
%where $x_j=1$ if vertex $j$ in one set and $x_j=0$ if vertex $j$ in the other set. The expression $(x_{i}+x_{j}-2x_{i}x_{j})$ evaluates to $1$ only when vertex $i$ and vertex $j$ belong to different sets; otherwise it evaluates $0$. Consequently, the ground states associated with the lowest energy will encode the maximum cut for the graph $G$.

%In this context, we have transformed the maximum cut problem into a QUBO formalism, which can be encoded according to the procedure outlined in the section dedicated to encoding QUBO problems. Thus we complete the encoding process. And the encoding cost for the entire process, including further encoding for the corresponding QUBO problem, is $O(n^2)$

\subsection{Dominating Set}

Given a graph $G=(V, E)$, with vertices $v_1,\dots,v_n$ and edges $e_1,\dots,e_m$. A dominating set $D$ is defined as a subset of vertices $D\subseteq V$ such that, for every vertex $v_i$ in the graph, either $v_i$ belongs to $D$ or $\exists v_j$ such that $(v_i,v_j) \in E$. The Dominating Set problem aims to minimize the size of $D$, which has been proved to be an NP-hard problem \cite{lewis1983michael}.

For each vertex $v_i$ we introduce
\begin{equation}
    a_i = \sqrt{\alpha_i}+\sum_{j:v_i\in e_j}\sqrt{\alpha_{n+j}}+\sqrt{\alpha_{n+m+1}},
\end{equation}
and for each edge $e_j$ we introduce two type numbers as follows
\begin{align}
    b_j &= -2\sqrt{\alpha_{n+j}},\\
    c_j &= -\sqrt{\alpha_{n+j}} + \sum_{i:v_i\in e_j}\sqrt{\alpha_i}.
\end{align}
To complete the encoding process, we additionally include 
\begin{equation}
    d_{i,n} = -n\sqrt{\alpha_i},
\end{equation}
for each vertex $v_i$, where $n\in[1,\text{degree}(v_i)]$.
Then target sum is 
\begin{equation}
    T = \sum_{i}\sqrt{\alpha_i}+k\sqrt{\alpha_{n+m+1}}.
\end{equation}

First, we will prove that, if graph $G$ has a dominating set $D$ of size $k$, the target sum can be satisfied. Suppose that the graph $G$ has a dominating set $D=\{v_{D_{1}},\dots,v_{D_{k}}\}$. If the vertex $v_i\in D$, we will include $a_i$ in the number subset. Consequently, the intermediate sum is 
\begin{equation}
    \sum_{i:v_i\in D}\sqrt{\alpha_{i}}+\sum_{i:v_i\in D}\sum_{j:v_i\in e_j}\sqrt{\alpha_{n+j}}+k\sqrt{\alpha_{n+m+1}}.
\end{equation}
The coefficient of $\sqrt{\alpha_{n+j}}$ will be 1 if only one vertex of $e_j$ belongs to the dominating set $D$, and 2 if both vertices of $e_j$ belong to $D$. In these two cases, we can add $b_j$ and $c_j$ respectively to eliminate the term $\sqrt{\alpha_{n+j}}$, where $j\in[1,m]$. This adjustment also ensures that the coefficient of each $\sqrt{\alpha_i}$ is  within $[1, \text{degree}(v_i)]$. Finally, we add appropriate $d_{i,n}$ for each vertex to eliminate the redundant $\sqrt{\alpha_i}$. Thus we achieve the target sum $T = \sum_{i}\sqrt{\alpha_i}+k\sqrt{\alpha_{n+m+1}}$.

Conversely, if we have a number subset that reaches the target sum, we can construct the dominating set based on its solution. Since the target sum $T$ contains $k\sqrt{\alpha_{n+m+1}}$, which can only comes from the numbers belong to $\{a_i\}$. Therefore, we first construct a subset $S$ with $k$ numbers associated with vertices, denoted as $\{a_{s(1)}, \dots, a_{s(i)}, \dots, a_{s(k)}\} $. Sum over subset $S$, we can get the intermediate sum,
\begin{equation}
    \sum_{i:v_i\in S}\sqrt{\alpha_{i}}+\sum_{i:v_i\in S}\sum_{j:v_i\in e_j}\sqrt{\alpha_{n+j}}+k\sqrt{\alpha_{n+m+1}}.
\end{equation}
As discussed above, the coefficient of $\sqrt{\alpha_{n+j}}$ will be 1 if only one vertex of $e_j$ belongs to $D$, and 2 if both vertices of $e_j$ belong to $D$. We also add $b_j$ and $c_j$ respectively to eliminate the number $\sqrt{\alpha_{n+j}}$. When adding $b_j$, no $\sqrt{\alpha_i}$ beyond the set $S$ is involved in the target sum. While adding $c_j$, terms of $\sqrt{\alpha_i}$ outside the subset $S$ are involved in the target sum. Since the target sum contains each $\sqrt{\alpha_i}$, it indicates that each vertex is either in the vertex set $S$ or is adjacent to the vertices in $S$. By definition, the vertex set is precisely a dominating set. And the rest $\sqrt{\alpha_i}$ terms can be eliminated by the appropriate $d_{i,n}$. Thus, we complete the proof. The encoding cost is 
\begin{equation}
    \text{Encoding cost}= n+4m.
\end{equation}

%Formally, if we identify a dominating set $D$ of graph G, then $V=\bigcup_{i=1}^{k} U_{D_i}$, where $U_{D_i}=\{v_{D_i},\mathcal{N}_{D_i}\}$, and $\mathcal{N}_{D_i}$ denotes the set of neighbor vertices of $v_{D_i}$. Therefore, we can encode it into a QUBO formula, inspired by the method discussed in the section on binary integer linear programming. We employ a binary variable $x_{i}$ to indicate whether vertex $v_i$ is included in the dominating set, where $x_i=1$ if vertex $v_i$ is included and $x_i=0$ otherwise. Additionally, we introduce auxiliary variables $y_{i,l}$ which count the occurrences of  vertex $v_i$ in the set $\{ U_{D_1},\dots U_{D_k}\}$. If $v_i$ appears $l$ times, then $y_{i,l}$ is 1, otherwise 0. Then the Hamiltonian is
%\begin{equation}
 %  H = \sum_{i=1}^{n}( 1-\sum_{l=1}^{n}y_{i,l} )^2+\sum_{i=1}^{n}(\sum_{l=1}^{n} ly_{i,l}-(x_i+\sum_{j:(i,j)\in E}x_j))^2+(k-\sum_{i=1}^{n}x_i)^2
%\end{equation}

%The first term ensures that the occurrences of vertex $v_i$ can take an exact number. The second term satisfies the definition of a dominating set. Finally, the last term guarantees that the size of the dominating set is exactly $k$. Due to the introduction of $n^2$ auxiliary elements $\{y_{i,l}\}$, the total encoding cost is $O(n^4)$.

\subsection{3-Coloring Problem}
%Given an undirected graph $G=(V, E)$, the 3-Coloring Problem aims to answer whether it is possible to color the vertices with, at most, three distinct colors, to make sure that no two adjacent vertices possess the same color.

Given an undirected graph $G=(V, E)$, a 3-coloring of $G$ is a function $\chi:V \to \{ red, green, blue \}$, such that for every edge $(u,v)\in E$, $\chi(u)$ and $\chi(v)$ are distinct. Hence, The objective of the 3-coloring problem is to determine whether there exists a 3-coloring for the given graph $G$, which is an NP-complete problem \cite{karp2010reducibility}.

For each vertex $i$, we introduce the numbers associated with them below,
\begin{align}
    p_{i}^{red}&=v_{i}+\sum_{j:i\in e_j}e_{j}^{red}=\sqrt{\alpha_{i}}+\sum_{j:i\in e_j}\sqrt{\alpha_{n+3j-2}},\\
    p_{i}^{green}&=v_{i}+\sum_{j:i\in e_j}e_{j}^{green}=\sqrt{\alpha_{i}}+\sum_{j:i\in e_j}\sqrt{\alpha_{n+3j-1}},\\
    p_{i}^{blue}&=v_{i}+\sum_{j:i\in e_j}e_{j}^{blue}=\sqrt{\alpha_{i}}+\sum_{j:i\in e_j}\sqrt{\alpha_{n+3j}},
\end{align}
where the numbers $\{v_i\}$ are utilized to differentiate between distinct vertices. Upon summing the numbers associated with neighboring sets $\{e_j^c | j \in \mathcal{N}(i)\}$, the resultant summation indicates that vertex $i$ is assigned the color $c$.

To complete the encoding procedure, additionally auxiliary numbers are included in set $S$ corresponding to each edge $j$,
\begin{align}
    e_{j}^{red}&=\sqrt{\alpha_{n+3j-2}},\\
    e_{j}^{green}&=\sqrt{\alpha_{n+3j-1}},\\
    e_{j}^{blue}&=\sqrt{\alpha_{n+3j}}.
\end{align}
Finally, we can set the sum target as
\begin{equation}
    T=\sum_{i}^{n}v_{i}+\sum_{j=1}^{m}(e_{j}^{red}+e_{j}^{green}+e_{j}^{blue})=\sum_{k=1}^{n+3m}\sqrt{\alpha_{k}}.
\end{equation}

Now we argue that the sum target can be reached is equivalent to graph $G$ has a 3-Coloring. Suppose that $G$ has a 3-Coloring. We construct a subset of $S$ as follows. For each vertex $i$, let subset $S'$ contain the number $p_{i}^{\chi(i)}$. At this stage, we have obtained the intermediate sum as 
\begin{equation}
    \sum_{i=1}^{n}v_{i}+\sum_{i=1}^{n}\sum_{j:i\in e_j}e_{j}^{\chi(i)}=\sum_{i=1}^{n}v_{i}+\sum_{j=1}^{m}\sum_{i:i\in e_j}e_{j}^{\chi(i)}.
\end{equation}
 Since each vertex in edge $j$ has a different color, we can reach the target sum by putting $e_{j}^{c}$ into subset $S'$, where $c \neq \chi(i)$, $i$ represent vertexes on edge $j$.

Conversely, suppose $S'$ is the subset whose sum equals $T$. Since each $v_{i}$ is counted exactly once in the target sum, it indicates that the number $p_{i}^{c}$ must be in the subset $S'$. let $c=\chi(i)$, then we will claim that $\chi$ is a valid coloring, i.e. that if $(u,v)\in E$ then $\chi(u) \neq \chi(v)$. Up to this step, subset $S'$ contain only $p_{i}^{\chi(i)}$ for each vertex $i$. Because we assume $\chi$ is a valid coloring map function, we just need to put proper $e_{j}^{c}$ into subset $S'$ to satisfy the target sum $T$. The construction of subset $S'$ can be directed to find the proper coloring for the original graph $G$. Therefore, we have proof that a graph $G$ has a 3-coloring if the correspondent subset sum problem is solvable. The encoding atom number is 
\begin{equation}
    \text{Encoding cost}=3(n+m).
\end{equation}

\section{Encoding Problems via QUBO}\label{subsec2}

Problems in this section cannot be reduced to subset sum problems directly, thereby implying the absence of an encoding scheme within the linear overhead. Unlike problems in section 3 completing the encoding process by reducing to subset sum problem directly. Here we introduce the Quadratic Unconstrained Binary Optimization (QUBO) problem as an intermediate reduction step. Initially, we elucidate the encoding approach for QUBO. Consequently, the problems in this section can be encoded by employing reduction to QUBO.
\subsection{Unconstrained Binary Optimization Problem}
The Quadratic Unconstrained Binary Optimization
(QUBO) problem is a well-known combinatorial optimization problem with a wide range of applications from finance and economics to machine learning 
~\cite{kochenberger2014unconstrained}. QUBO problem is to find the optimal binary vector that minimizes a quadratic objective function subject to no constraints on the binary variables. More formally, given a matrix Q, a function with a binary vector $b=b_{1},\dots,b_{n}$ is defined as $f_{Q}(x)=b^{T}Qb=\sum_{i=1}^{n}\sum_{j=1}^{n}Q_{ij}b_{i}b_{j}$.

QUBO problem is also closely related and equivalent to the Ising model, according to a standard process we can transform the objective function into Ising form as
\begin{equation}
    H(S)=-\sum_{\langle i,j \rangle}J_{ij}s_{i}s_{j}-\mu \sum_{j}h_{j}s_{j}.
\end{equation}
In consideration of hardware constraints, where qubits' interactions are quasi-local within realistic physical platforms, the development of efficient encoding schemes for achieving all-to-all coupling in atom-cavity systems becomes crucial. To address this, we adopt the method discussed in \cite{lechner2015quantum, glaetzle2017coherent, ye2023universal}, which involves the introduction of parity code $z_{ii'}=s_{i}s_{i'}$. The interactions in $J$ then become local fields with a cost of introducing constraints $z_{ii'}z_{ii'+1}z_{i+1i'+1}z_{i+1i'}=1$. The total number of these independent constraints is $(n-1)(n-2)/2$. Then we rewrite the constraints in terms of the 3-SAT formula,
\begin{equation}
    \begin{aligned}
    &(\beta \lor x_{ii'} \lor x_{ii'+1})
	\land(\beta\lor\bar{x}_{ii'}\lor\bar{x}_{ii'+1}) 
\land(\beta \lor x _{i+1i'+1}\lor x_{i+1i'})\\
&\land(\beta \lor\bar{x}_{i+1i'+1}\lor\bar{x}_{i+1i'})
\land(\bar{\beta} \lor\bar{x}_{ii'}\lor x_{ii'+1}) 
	\land(\bar{\beta} \lor x_{ii'}\lor\bar{x}_{ii'+1})\\
&\land(\bar{\beta} \lor\bar{x}_{i+1i'+1}\lor x_{i+1i'})
	\land(\bar{\beta} \lor x_{i+1i'+1}\lor\bar{x}_{i+1i'})
    \end{aligned}
\end{equation}
with $x_{ii'}=(z_{ii'}+1)/2$, and $\beta$ an introduced auxiliary variable. Taking all constraints into account, we have a 3-SAT problem with $n'=(n-1)^2$ variables, and $m'=4(n-1)(n-2)$ clauses. 

As shown in the previous section, the 3-SAT formula is equivalent to the subset sum problem. Therefore, the QUBO problem becomes optimizing
\begin{equation}
    {\rm min}_{} \left\{
\sum_{k=1}^{n(n-1)/2} J_{k} y_k  \right. \\ 
 \left.  + \Delta \left(  \sum_{k=1} ^{2(n-1)(5n-9) }  y_k r_k (I_{\rm const}, B_{\rm const} )^2    \right) ^2 
			\right\},
\end{equation}
with the first $n(n-1)/2$ elements of $y_{k}$ representing $x_{ii'}$, and $J_{k}$ representing $J_{ii'}$ correspondingly. Here energy penalty term $\Delta>0$ should be sufficiently large to enforce the required constraints. According to our method, the overhead in the cost of atom number for the QUBO problem is $2(n-1)(5n-9)-n\sim O(n^2)$, which is quadratic scaling of problem size.

In this section, we construct an explicit and straightforward procedure for encoding the QUBO problem. This approach offers a systematic method whereby any combinatorial optimization problem can be encoded within the atom-cavity system, provided they possess corresponding QUBO formulas. Subsequently, the following problems are all encoded by first mapping them to QUBO formulas.

\subsection{Binary Integer Linear Programming}

%The fundamental formulation of a Binary Integer Linear Programming (BILP) problem originates from the optimization of a linear objective function under a collection of linear constraints. A defining characteristic of BILP problems is the constraint that the decision variables must assume binary values.

The binary integer linear programming is to optimize the objective function $c^{T}x$ under the constraint $Ax=b$, where $x$ is a vector with binary variables $x_1,x_2,\dots,x_n$. It is natural to give a corresponding optimization function as
\begin{equation}
    H=\alpha\sum_{j=1}^{m}\lbrack b_j-\sum_{i=1}^{n}A_{ji}x_i\rbrack^2-\beta\sum_{i=1}^{n}c_ix_i.
\end{equation}
The term $\alpha\sum_{j=1}^{m}\lbrack b_j-\sum_{i=1}^{N}A_{ji}x_i\rbrack^2$ penalizes solutions that do not satisfy the constraint $Ax=b$, while the term $-\beta\sum_{i=1}^{n}c_ix_i$ facilitates the identification of the optimal solution within the programming problem.

To ensure that the constraints are not violated, it is necessary to determine the bound on the ratio  $\alpha/\beta$. Assume that the constraint $Ax=b$ can be satisfied for some assignment of $x$. Suppose a specific variable $x_i$ is flipped, resulting in an energy change of $\alpha\sum_{j}^{m}A_{ji}^2-\beta c_i+2\beta c_ix_i$. We require this change to be greater than zero. Consequently, we obtain the inequality $\frac{\alpha}{\beta}>\frac{c_i(1-2x_i)}{\sum_{j}^{m}A_{ji}^2}$. Therefore, to ensure this condition is satisfiable for any possible case, we can set $\frac{\alpha}{\beta}>\frac{max(c_i)}{min(\sum_{j}^{m}A_{ji}^2})$, and the encoding cost is $O(n^2)$.

\subsection{Knapsack with Integer Weights}

In the knapsack problem, it is given $n$ items with weight $w_1,w_2,\dots,w_n$ and value $v_1,v_2,\dots,v_n$, and a Knapsack with a specified capacity $W$. The knapsack problem is aimed to maximize the total value $\sum_{i=1}^{n} v_ix_i$ without surpassing its weight limit $\sum_{i=1}^{n} w_ix_i\leq W$. 

We now derive the first QUBO reformulation of the knapsack problem as considered in~\cite{lucas2014ising}. To deal with the inequality for problem constraint, here introduce additional binary variables $y_1,\dots,y_W$ in which $y_k$ is 1 if the final weight of the knapsack is $k$, and 0 otherwise. Consequently, the weight limit constraint can be encoded as follows:
\begin{equation}
    H_{constraint}=(1-\sum_{k=1}^{W}y_k)^2+(\sum_{k=1}^{W}ky_k-\sum_{i=1}^{n} w_ix_i)^2,
\end{equation}
which makes sure that all possible weight limits can be reached. Finally we add optimization term $H_{opt}=-\Delta \sum_{i=1}^{n}v_ix_i$ to maximize the objective function.
Then the entire Hamiltonian is 
\begin{equation}
    H = (1-\sum_{k=1}^{W}y_k)^2+(\sum_{k=1}^{W}ky_k-\sum_{i=1}^{n} w_ix_i)^2-\Delta \sum_{i=1}^{n}v_ix_i.
\end{equation}

To ensure that no solution allows $H_{constraint}$ to be violated in favor of making $H_{opt}$ more negative, it is necessary to select $\Delta$ sufficiently small. Specifically, if $H_{constraint}$ is violated once by flipping one spin,  the resulting energy penalty is $w_i^2$, while the change in energy contributed by the optimization term is $-v_i\Delta$. Consequently, we derive that $\Delta \leq \frac{w_i^2}{v_i}$. Extending to the general case in this Hamiltonian, this condition can be further refined to $\Delta \leq \frac{1}{max(v_i)}$.
Finally, the encoding cost becomes  $O((n+W)^2)$.

\subsection{Job Sequencing Problem}

The job-shop sequencing problem involves the scheduling of a set of independent jobs $J=\lbrace J_1, J_2,\dots, J_n\rbrace$ on $m$ machines, each characterized by an integer processing time $t_i$ and a deadline $T_0$ for completion. The central objective is to determine an optimal distribution for the jobs such that deadline $T_0$ can be satisfied. The encoding method for the job sequencing problem to a QUBO formula has been discussed in \cite{lucas2014ising}.

The Hamiltonian is presented below
\begin{equation}
    H=\sum_{i=1}^{n}(1-\sum_{j=1}^{m}x_{i,j})^2+\sum_{j=1}^{m}(1-\sum_{k=1}^{T_0}y_{k,j})^2+\sum_{j=1}^{m}(\sum_{k=1}^{T_0}ky_{k,j}-\sum_{i=1}^{n}t_ix_{i,j})^2,
\end{equation}
where $x_{i,j}$ denotes the $i$-th job $J_i$ are given to the $j$-th machine, and {$y_{k,j}$} are auxiliary variables which denote the potential time consumption for the $j$-th machine finishing its jobs ranging from $1$ to $T_0$. Now we look back to the Hamiltonian and explain each term respectively. The first term confirms that each job is divided into exactly one machine, and the second term makes sure the ground state chooses a special auxiliary variable value $y_{k,j}$ which indicates the total time consumption is $k$. The last term is aimed to satisfy the demand of deadline $T_0$, punishing the situation that any machine consumption time is larger than $T_0$.
The final component of the Hamiltonian is designed to meet the time deadline, which will penalize instances wherein any given machine exceeds the time threshold $T_0$. And the corresponding encoding atom number is $O(m^2(n+T_0)^2)$.

\subsection{Directed Hamiltonian Cycle}
A Hamiltonian cycle in a directed graph $G(V, E)$ is a directed cycle $C$ that includes every vertex in the set $V$ exactly once. Formally, it is a sequence of vertices $v_{i_{1}}, v_{i_{2}}, \dots, v_{i_{n}}, v_{i_{1}}$, where $n$ is the number of vertices, satisfying the following conditions:$(v_{i_{k}},v_{i_{k+1}})\in E$ for $\forall k\in \lbrack 1,n)$, and additionally $(v_{i_{n}},v_{i_{1}}) \in E$. The QUBO formula has been previously derived in \cite{lucas2014ising}. In this section, we will provide a review of this formulation.

Here we can introduce $n^2$ binary variables $x_{v, i}$ to formulate the problem Hamiltonian. Variable $x_{v, i}$ denotes that the vertex $v$ occupies the $i$-th position in the Hamiltonian cycle. The corresponding Hamiltonian is represented below

\begin{equation}
    H=\sum_{v=1}^{n}(1-\sum_{i=1}^{n}x_{v,i})^2+\sum_{i=1}^{n} (1-\sum_{v=1}^{n}x_{v,i})^2+\sum_{(uv\notin E)}\sum_{i=1}^{n}x_{u,i}x_{v,i+1}.
\end{equation}

It is important to note that in the final term, the summation index $(uv)$ is directed, implying that the order of $uv$ is significant. For undirected graphs, edges $(vu)$ and $(uv)$ are added to the edge set $E$, allowing the graph to be treated as a directed graph. The first term ensures that each vertex $v$ in the graph appears only once in the cycle, while the second term guarantees that each step of the cycle contains exactly one vertex. The final term penalizes pairs of sequentially adjacent vertices in the cycle that do not correspond to edges in the set $E$ of the graph $G$. Therefore, the ground states of this Hamiltonian represent the solutions to the Hamiltonian cycle problem. The encoding cost, in terms of the number of atoms required, is $O(n^4)$.

\subsection{Traveling Salesman Problem}
 Formally, the traveling salesman problem can be modeled as a complete weighted graph $G=(V, E, W)$, where $V$ denotes the set of vertices, $E$ denotes the set of edges, and $W(u,v)$ represents the weight associated with each edge $(u,v)$, corresponding to the distance between vertices $u$ and $v$. Thus, the traveling salesman problem aims to identify a Hamiltonian cycle with the minimum possible total weight.

 Next, we present the problem Hamiltonian by first providing the QUBO formulation of the traveling salesman problem as outlined in \cite{lucas2014ising}. According to the definition of the traveling salesman problem, the problem Hamiltonian of the Hamiltonian cycle serves as a constraint Hamiltonian for the traveling salesman problem. After combined with the optimization term  $\Delta\sum_{(uv)\in E)}W_{uv}\sum_{j=1}^{n}x_{u,j}x_{v,j+1}$, the complete Hamiltonian is given by

\begin{equation}
\begin{split}
    H =& \sum_{v=1}^{n}(1-\sum_{j=1}^{n}x_{v,j})^2+\sum_{j=1}^{n}(1-\sum_{v=1}^{n}x_{v,j})^2\\
    &+\Delta\sum_{(uv)\in E}W_{uv}\sum_{j=1}^{n}x_{u,j}x_{v,j+1}.
\end{split}
\end{equation}
It is important to note that in the traveling salesman problem, since the graph $G$ is a complete graph, the term $\sum_{(uv \notin E)} \sum_{i=1}^{n} x_{u, i} x_{v,i+1}$ in the Hamiltonian cycle problem is removed.

Here we choose $\Delta$ small enough that it is never favorable to violate the constraints Hamiltonian correspond to the directed cycle. The determination of the bound for $\Delta$ is similar to that in the Knapsack problem. Considering the worst-case scenario, we require $\Delta \leq \frac{1}{\max(W_{uv})}$. The number of encoding atoms remains consistent with that of the directed Hamiltonian cycle, denoted as $O(n^4)$.

\section{Conclusion}
\label{sec4}

In this work, we have developed quantum computing encoding schemes for various NP-complete problems,  exploiting the potential of cold atoms within an optical cavity. Recent studies \cite{blais2021circuit, bentsen2020tunable, periwal2021programmable} highlight the effectiveness of organizing atoms inside the cavity, demonstrating their capability to participate in long-range interactions facilitated by four-photon processes coupled with cavity modes \cite{ye2023universal}. This characteristic makes them highly suitable for encoding subset sum problems. Building upon these insights, we propose an expansion of this quantum computing framework to tackle a broader spectrum of NP-complete problems.

Our approach involves the introduction of a series of auxiliary squarefree integers and the subsequent reduction of various combinatorial optimization problems to the subset sum problem formula, which can be seamlessly implemented in experimental atom-cavity systems. 
 
We show a number of NP-complete problems have efficient encoding schemes using the atom-cavity system, having a linear cost in the atom number. 
Our theory implies the atom-cavity system is a promising platform to demonstrate a practical advantage in solving various NP-complete problems. 
Nonetheless, there still remain certain problems such as QUBO, integer programming, and the Hamiltonian cycle 
for which the scaling in the atom number cost is quadratic or higher.  

It is worth remarking here that the encoding schemes we develop here may also be adopted in other optical systems for solving NP-complete problems, where a similar form of Mattis-type spin glass Hamiltonian as in the atom-cavity system can be implemented \cite{pierangeli2019large, pierangeli2020adiabatic, huang2021antiferromagnetic, sun2022quadrature, jacucci2022tunable,wang2024efficient}.

\section*{Declarations}

\begin{itemize}

\item \textbf{Acknowledgement} 

We acknowledge the helpful discussion with Zhengfeng Ji.

\item \textbf{Funding}

This work is supported by 
National Key Research and Development Program of China (2021YFA1400900), 
Innovation Program for Quantum Science and Technology of China (2024ZD0300100), 
National Natural Science Foundation of China (11934002), 
and Shanghai Municipal Science and Technology Major Project (Grant No. 2019SHZDZX01). 

\item \textbf{Competing interests} 

No competing interests.

\item \textbf{Data and materials availability  availability}

Data and materials will be made available on reasonable request.

\item \textbf{Authors' contributions}

Meng Ye did the theoretical proof under the supervision of Xiaopeng Li. All authors contributed to the discussions of the results
and the writing of the manuscript. 

\end{itemize}

\bibliography{references}% common bib file
%\input sn-article.bbl

\end{document}